\documentclass[12pt,preprint]{aastex}

\usepackage{color,natbib,pdflscape}
\shorttitle{Stochastic Variability in GRS 1915+105}
\shortauthors{Polyakov, Neilsen, \& Timashev}

\begin{document}


\title{STOCHASTIC VARIABILITY IN X-RAY EMISSION FROM THE BLACK HOLE BINARY GRS 1915+105}


\author{Yuriy S. Polyakov}
\affil{USPolyResearch, Ashland, PA 17921, USA}
\email{ypolyakov@uspolyresearch.com}

\author{Joseph Neilsen\altaffilmark{1}}
\affil{MIT Kavli Institute for Astrophysics and Space Research, Cambridge, MA 02139, USA}

\and

\author{Serge F. Timashev\altaffilmark{2}}
\affil{Karpov Institute of Physical Chemistry, Moscow 103064, Russia}

\altaffiltext{1}{Harvard-Smithsonian Center for Astrophysics, 
Harvard University, Cambridge, MA 02138, USA}

\altaffiltext{2}{Institute of Laser and Information Technologies, Russian Academy of Sciences, Troitsk,
Moscow Region 142190, Russia} 


\begin{abstract}
We examine stochastic variability in the dynamics of X-ray emission from the black hole system GRS 1915+105, a strongly variable microquasar commonly used for studying relativistic jets and the physics of black hole accretion. The analysis of sample observations for 13 different states in both soft (low) and hard (high) energy bands is performed by flicker-noise spectroscopy (FNS), a phenomenological time series analysis method operating on structure functions and power spectrum estimates. We find the values of FNS parameters, including the Hurst exponent, flicker-noise parameter, and characteristic time scales, for each observation based on multiple 2,500-second continuous data segments. We identify four modes of stochastic variability driven by dissipative processes that may be related to viscosity fluctuations in the accretion disk around the black hole: random (RN), power-law (1F), one-scale (1S), and two-scale (2S). The variability modes are generally the same in soft and hard energy bands of the same observation. We discuss the potential for future FNS studies of accreting black holes.
\end{abstract}


\keywords{accretion, accretion disks --- binaries: close --- black hole physics --- methods: data analysis --- methods: statistical --- X-rays: individual (GRS 1915+105)} 



\section{INTRODUCTION}

A microquasar is a binary system in which a normal star orbits a compact object, either a neutron star or a black hole with a typical mass of 10 solar masses. Under certain conditions, matter may be transferred from the normal star onto the compact object in the form of an accretion disk \citep[for theory, see][]{Sha73}. The subsequent release of gravitational potential energy can lead to the formation of relativistic jets \citep{Bla77}, as well as strong, variable emission across a broad range of wavelengths, from radio waves to X-ray and $\gamma$-ray \citep{Tan72,Mir94}. Due to their short characteristic time scales compared to supermassive black holes (i.e.,\ black holes having a typical mass of several million solar masses), microquasars are regarded as key astrophysical objects for studying relativistic jets and the physics of the accretion process \citep{Sam96,Mir99,McH06}. 

The diverse physical processes in microquasars may be studied indirectly by examining temporal variability in emission across the electromagnetic spectrum. Detailed analysis of temporal variations may be used to estimate the geometry and structure of a binary system and to test various astrophysical models for microquasars \citep{Bel10}. This variability is commonly characterized \citep[][and references therein]{K06} with estimates of the power spectrum of the light curve \citep{Bel02,Bel06,Qu10,Mac11,Nei11}, frequency-dependent time lags between different energy bands \citep{Rei00}, or a combination of total count rate and color-color diagrams \citep[colors are ratios of X-ray count rates in different energy bands;][]{Bel10}. In the first method, any apparent noise components and quasi-periodic oscillations (QPOs) in the power spectra are treated as superpositions of broken power laws and/or Lorentzians \citep{Bel02}. The second method is based on the analysis of phase lag spectra for ``soft" (lower) and ``hard" (higher) energy bands, which are believed to be related to two different sources of X-ray emission in the accretion flow; lags between energy bands can help constrain the emission geometry. The third method uses ratios between X-ray lightcurves in hard and soft energy bands as a proxy for studying the variability of the energy spectrum over time.

One of the essential features observed in many X-ray emitting astrophysical systems is strong apparent randomness on a range of time scales. While the methods above have proven successful in discerning and quantifying different variability modes, they could not elucidate the origin of the apparent randomness, which remains an open question. For instance, it may be related to variation of some external parameter, such as stochastic fluctuations of the viscosity or mass accretion rate in the accretion disk \citep{L97,T07,Wil09,Utt11}; some complex variability may be due to chaotic (deterministic) fluctuations produced by inner-disk instabilities \citep{Mis04,Mis06}. Other interpretations are possible, and a complete picture of energy-dependent variability has yet to emerge.  The origin of apparent randomness in each particular case can be studied using the methods of complexity science, such as deterministic chaos, multifractal theory, flicker-noise spectroscopy, fractional calculus, cellular automata theory, and other approaches. In this paper, we apply a phenomenological complexity science method to analyze one of the most remarkable astrophysical systems, the Galactic microquasar GRS 1915+105, which consists of a black hole with mass $14\pm4 M_\odot$, where $M_\odot$ is the mass of the Sun, in a 33.5-day orbit around a donor star with mass $1.2\pm0.2 M_\odot$ \citep{G01}. GRS 1915+105 is famous for its relativistic jets \citep{Mir94} and strong X-ray variability in 14 different states \citep{B00,K02,H05}, which are characterized by highly-structured low-frequency variability in the X-ray lightcurve on time scales ranging from seconds to hours, with amplitudes often well in excess of 100\% \citep{B00}. Accordingly, it provides a useful case for studying the underlying processes behind the X-ray variability.

Apparent random behavior in the X-ray lightcurves was examined for 12 of the states using nonlinear time series analysis techniques \citep{Mis04,Mis06,Har11}. It was suggested that three types of underlying mechanisms may take place: chaotic, stochastic (random), and ``nonstochastic" (chaotic + colored noise), which implies that several of the 12 variability states could be explained in terms of a deterministic system of ordinary nonlinear differential equations. The analysis presented by \citet{Mis04} was based on the estimation of correlation dimension, which is complicated in this case by the presence of high-frequency Poisson noise in all light curves. Moreover, because the correlation dimension can approach a constant value for both chaotic data and colored noise, it is possible that colored noise in the light curve data for GRS 1915+105 may be incorrectly classified as chaotic \citep{Osb89}. A more advanced procedure including surrogate data analysis, singular value decomposition (SVD) technique, and correlation dimension was applied by \citet{Mis06} in an attempt to obtain more conclusive results. The analysis of SVD plots for real and surrogate data revealed some apparent attractors  in the real X-ray variability of GRS 1915+105. However, contamination by noise made it difficult to determine whether the attractors were chaotic or periodic (i.e., limit cycles).
Correlation entropy and multifractal analysis were added to the nonlinear time series analysis of GRS 1915+105 data by \citet{Har11}. This allowed the authors to identify colored noise and estimate the range of fractal dimensions for each observation. It should be noted that in addition to the challenges mentioned above, the light curve data are generally nonstationary and may exhibit a range of characteristic time scales, which further complicates the application of nonlinear time series analysis to these data and makes the results inconclusive, as pointed out by \citet{Har11}. 

Some of these problems can be overcome by utilizing alternative methods of signal analysis for complex systems that are not based on chaos theory. In this study, we apply one such approach, flicker-noise spectroscopy (FNS) \citep{Tim10a,Tim12,Tim07a,Tim06a,Tim07b}, a phenomenological method operating on Kolmogorov transient structure functions and autocorrelation-based estimates of power spectrum, to examine the apparent randomness in the X-ray lightcurves of GRS 1915+105, and compare our findings with previously reported results obtained by nonlinear time series analysis \citep{Har11}. In contrast to chaos theory, the FNS method separates between regular and stochastic components, handles multiple characteristic time scales, and provides a mechanism for excluding the effect of Poisson noise at highest frequencies, thus addressing most of the challenges mentioned above.

\section{FLICKER-NOISE SPECTROSCOPY}

Here, we will only deal with the basic FNS relations needed to understand the parameterization procedure. FNS is described in more detail elsewhere \citep{Tim06a,Tim07a,Tim07b,Tim08a,Tim10a}.

It should be noted that the term ``stochastic", here and further in this paper, refers to random variability in the signals of complex systems characterized by nonlinear interactions, dissipation, and inertia \citep{Tim10a,Tim07a}. Conceptually, the method separates the analyzed signal into three components: low-frequency regular component corresponding to system-specific ``resonances"  and their interferential (nonlinear) contributions, stochastic ``jump" (random walk) component at larger frequencies corresponding to a dissipative process of anomalous diffusion, and stochastic highest-frequency ``spike" component corresponding to inertial (non-dissipative) effects. 
The idea of separating the stochastic dynamics of a complex nonlinear system into dissipative random-walk and non-dissipative inertial components stems from numerous simulations performed by various cellular automata models demonstrating that the stochastic dynamics of these processes is associated with intermittency, consecutive alternation of rapid (inertial) changes in the values of dynamic variables on small time intervals with small (dissipative) variations of the values on longer time intervals \citep{Bak97,Wol02, Sch95}. These studies showed that the occurrence of dissipative and inertial changes in such dynamics can be responsible for the power law distributions observed in nature, such as Guttenberg-Richter and Zipf-Parreto laws and flicker noise. The FNS method further assumes that correlations present in sequences of different irregularities, such as spikes and jumps, may be treated as main information carriers in the stochastic 
variability of a complex dissipative system \citep{Tim07a}. This assumption has much in common with how wavelet analysis evaluates the inverse transform only for those time intervals where the first derivative of the signal experiences steepest changes, i.e., intervals with the highest degree of irregular behavior \citep{Wal02}. The spike-like irregularities are naturally associated in FNS with the inertial component, and jump-like irregularities with a dissipative process of anomalous diffusion (random walks), which is illustrated in Fig. 2 of \cite{Tim07a}. It should be pointed out that in practical signal analysis, Kolmogorov transient structure functions and power spectrum estimates contain only integral contributions from stochastic irregularities corresponding to a specific evolution hierarchy level and distributed with certain probability densities. If one analyzes the signal at a much higher sampling frequency, each of the irregularities observed at a smaller sampling frequency becomes a complex structure with its own spike- and jump-like irregularities. Further differences and similarities between FNS and other methods for the analysis of the dynamics of complex systems, such as nonlinear dynamics (deterministic chaos), wavelet analysis, and fractional calculus, are discussed in depth elsewhere \citep{Tim07a}. 

As FNS is a phenomenological method, its main goal is to estimate from analyzed series certain model-independent stochastic parameters corresponding to general dissipative and inertial processes. These parameters can then be used to develop specific models or refine existing ones. The validity of FNS assumptions for many applications was proven by effectively applying the method to the analysis of structure and dynamics for various electrochemical, geophysical, medical, and physical systems \citep{Des03,Tel04,Hay06,Tim06a, Ida07,Tim07a,Tim09,Tim10a,Tim10b,Rya11,Mir11,Tim12}. For instance, it was recently demonstrated by FNS that the dynamics of X-ray emission from GRS 1915+105 and Cygnus X-1, recorded in the timeframe from January 1, 1996 to December 31, 2005 with a sampling interval of approximately 1.5 hrs, manifested a one-scale process of anomalous diffusion \citep{Tim10a}. Therefore, it would be logical to apply the FNS method to the analysis of the GRS 1915+105 X-ray emission data at much smaller (subsecond) sampling intervals, which is the topic of this study.

In FNS, all introduced parameters for signal $V(t)$, where $t$ is time, are related to the autocorrelation function
\begin{equation}
\psi (\tau) = \left\langle {V(t)V(t + \tau )} \right\rangle_{T-\tau}, \label{eq1}
\end{equation}
where $\tau$ is the time lag parameter ($0 < \tau \le T_M$), $T_M$ is the upper bound for $\tau$ ($T_M \le T/2$), and $T$ is the averaging window. The angular brackets in relation (\ref{eq1}) stand for the averaging over time interval $T-\tau$:  
\begin{equation}
\left\langle {(...)} \right\rangle_{T-\tau}  = {1 \over {T-\tau}}\int^{T-\tau}_{0} {(...) \,dt}. \label{eq2}
\end{equation} 
The averaging over interval $T-\tau$ implies that all the characteristics that can be extracted by analyzing functions $\psi(\tau)$ should be regarded as the average values on this interval.  

To extract the information contained in $\psi (\tau )$ ($\left\langle {V(t)} \right\rangle = 0$  is assumed), the following transforms, or ``projections", of this 
function are analyzed: cosine transforms (power spectrum estimates) $S(f)$, where $f$ is the frequency,
\begin{equation}
S(f) = 2 \int^{T_M}_{0} { \left\langle {V(t)V(t + t_1 )} \right\rangle_{T-\tau} \, \cos({2 \pi f t_1}) \,dt_1} \label{eq3}
\end{equation}
\newline and its difference moments (Kolmogorov transient structure functions) of the second order $\Phi^{(2)} (\tau)$
\begin{equation}
\Phi^{(2)} (\tau) = \left\langle {\left[ {V(t) - V(t+\tau )} \right]^2 } \right\rangle_{T-\tau}. \label{eq4}
\end{equation}

The information contents of $S(f)$ and $\Phi^{(2)}(\tau)$ are generally different, and the parameters for both functions are needed to solve parameterization problems. By considering the intermittent character of signals under study, interpolation expressions for the stochastic components $S_s(f)$ and ${\Phi_s}^{(2)} (\tau)$ of $S(f)$ and $\Phi^{(2)} (\tau)$, respectively, were derived using the theory of generalized functions by \citet{Tim06a}. It was shown that the stochastic components of structure functions $\Phi^{(2)} (\tau)$ are formed only by jump-like (random walk) irregularities, and stochastic components of functions $S(f)$, which characterize the ``energy side" of the process, are formed by spike-like (inertial) and jump-like irregularities. It should be noted that $\tau$ in Equations (\ref{eq1})-(\ref{eq4}) is considered as a macroscopic parameter exceeding the sampling period by at least one order of magnitude. This constraint is required to derive the expressions and separate out contributions of dissipative jump-like and inertial (non-dissipative) spike-like components.

The basic idea in parameterization is to use two interpolation expressions for the stochastic components. The first 
 expression is used to determine the spectral contribution of the stochastic components of signal $V(t)$ and exclude the contribution of the low-frequency regular component to the parameters related to jump- and spike-like stochastic irregularities. The second interpolation, which deals  with the stochastic component in the structure function, is used to determine the parameters characterizing the series of jump-like  irregularities, which correspond to anomalous diffusion \citep{Tim10a}.

For the case of one characteristic scale in the sequences of jump-like and spike-like irregularities, the spectral interpolations for both types of stochastic irregularities have the following functional form \citep{Tim06a}:

\begin{equation}
S_{s} (f) \approx {{S_{s} (0)} \over {1 + (2\pi f T_{0} )^{n} }},
\label{eq5}
\end{equation}
where $S_s(0)$,  $T_0$, and $n$ and are phenomenological parameters, which have different meanings for spike- and jump-like irregularities. In other words, the resultant interpolation is a superposition of expressions (\ref{eq5}) for these two components. When the contributions of spikes and jumps to the overall stochastic component are comparable in the highest-frequency range or the contribution of spikes is negligible, which is true for the data analyzed in this study, Equation (\ref{eq5}) can be readily used as the resultant interpolation. In this case, $S_{s} (0)$ is the phenomenological parameter corresponding to the initial (plateau) value of power spectrum estimate, $T_{0}$ is the correlation time for jump- and spike-like irregularities after which the self-similarity observed in power spectrum estimate breaks down, and $n$ is the flicker-noise parameter characterizing the rate of loss of correlations in the series of high-frequency irregularities in time intervals $T_{0}$. It should be noted that the spectral interpolations for spike- and jump-like irregularities were derived using the theory of generalized functions based on different assumptions \citep{Tim06a}, and generally correspond to different physical processes. The negligible contribution of the spike component in this particular case is attributed to the high level of Poisson noise at highest frequencies, as will be illustrated later in the paper. It should also be pointed out that mathematically Equation (\ref{eq5}) represents a superposition of flat spectrum at lower frequencies with power law at higher frequencies (broken power law with a smooth transition), and $T_{0}$ is related to the reciprocal of the characteristic frequency used in astrophysical spectral analysis \citep{Bel02}.

The structure function interpolation for the case of one characteristic scale in the sequence of jump-like irregularities (random walks) has the following form \citep{Tim06a}:

\begin{equation}
\Phi_s ^{(2)} (\tau ) \approx 2\sigma ^2 \left[ {1 - \Gamma \,^{ - 1} (H) \cdot \Gamma \,(H ,\tau /T_1 )} \right]^{_2 }, \label{eq6}
\end{equation}
where $\Gamma (s, x) = \int\limits_x^\infty  {\exp(-t) \cdot t^{s - 1}} dt$ and $\Gamma (s) = \Gamma (s, 0)$ are the complete and incomplete gamma functions, respectively ($x \ge 0$ and $s > 0$); $\sigma$ is the standard deviation of the measured dynamic variable with dimension [$V$]; $H$ is the Hurst exponent, which describes the rate at which the dynamic variable ``forgets" its values on the time intervals that are less than the correlation time $T_1$. At $H = 0.5$, interpolation (\ref{eq6}) corresponds to Brownian motion (Fickian diffusion); at $H < 0.5$, to subdiffusion; and at  $H > 0.5$, to superdiffusion (L\'{e}vy diffusion or L\'{e}vy flights) \citep{Tim10a}. Subdiffusion and superdiffusion are nonlinear stochastic processes that are generally modeled by fractional Fokker-Planck equations \citep{Tsa02,Che10}. Expression (\ref{eq6}) at $H=0.5$ was shown to correspond to the diffusion equation with integrodifferential boundary conditions \citep{Tim10a}. 


For the case of two characteristic scales, the resultant spectral interpolation can be written as
\begin{equation}
S_{s} (f) \approx {{{S_{s1} (0)} \over {1 + (2\pi f T_{01} )^{n1} }} +{ {S_{s2} (0)} \over {1 + (2\pi f T_{02} )^{n2} }}},
\label{eq7}
\end{equation}
where indices 1 and 2 in subscripts denote respective scales \citep{Par00}.

The structure function for two characteristic scales takes the following piecewise form:
\begin{equation}
\Phi_s ^{(2)} (\tau ) \approx 2 {\sigma_1}^2 \left[ 1 - {{\Gamma (H_1 ,\tau /T_{11} )} \over {\Gamma (H_1 ) } } \right]^{2 }  ,\,\,\,\, {\tau < \tau_0}, \label{eq8}
\end{equation}
\begin{equation}
\Phi_s ^{(2)} (\tau ) \approx 2 {\sigma_1}^2 + 2 {\sigma_2}^2 \left[ 1 - { {\Gamma (H_2,\frac {\tau-\tau_0} {T_{12}} )} \over {\Gamma (H_2 ) } } \right]^{2 }  ,\, {\tau \ge \tau_0}, \label{eq9}
\end{equation}
where indices 1 and 2 in subscripts denote respective scales, and $\tau_0$ is the value of $\tau$ corresponding to the transition to the second scale. Expressions (\ref{eq8}-\ref{eq9}) were derived based on the assumption that the characteristic scales are much apart from each other, which justifies the use of a linear superposition instead of the more complex expression employed by \citet{Par00}.

In some applications, it is also useful to introduce parameters $S_s(T_0^{-1})$ (for one-scale case) and $S_s(T_{01}^{-1})$, $S_s(T_{02}^{-1})$ (for two-scale case) to account for the ``intensity" of jump- and spike-like irregularities on characteristic frequency intervals \citep{Tim12}.

FNS interpolations were derived for a wide-sense stationary signal in which the phenomenological parameters are the same at each level of 
 the system evolution hierarchy. At the same time, they can also be applied to the analysis of real signals, which are generally 
  nonstationary, but can be characterized by a standard deviation within a specific averaging window. In this case, the real signals at specific averaging intervals 
and sampling frequencies should be regarded as 
 quasi-stationary with certain values of standard deviation and other phenomenological parameters. It should be noted that the values of phenomenological parameters may vary on 
 different quasi-stationary intervals.

The detailed FNS parameterization algorithms for one- and two-scale cases in discrete form, which were adapted for the X-ray emission data, are described in Appendices A and B, respectively.

\section{RESULTS AND DISCUSSION}

 \begin{figure}
\begin{center}
\figurenum{1}
\includegraphics[width=15cm]{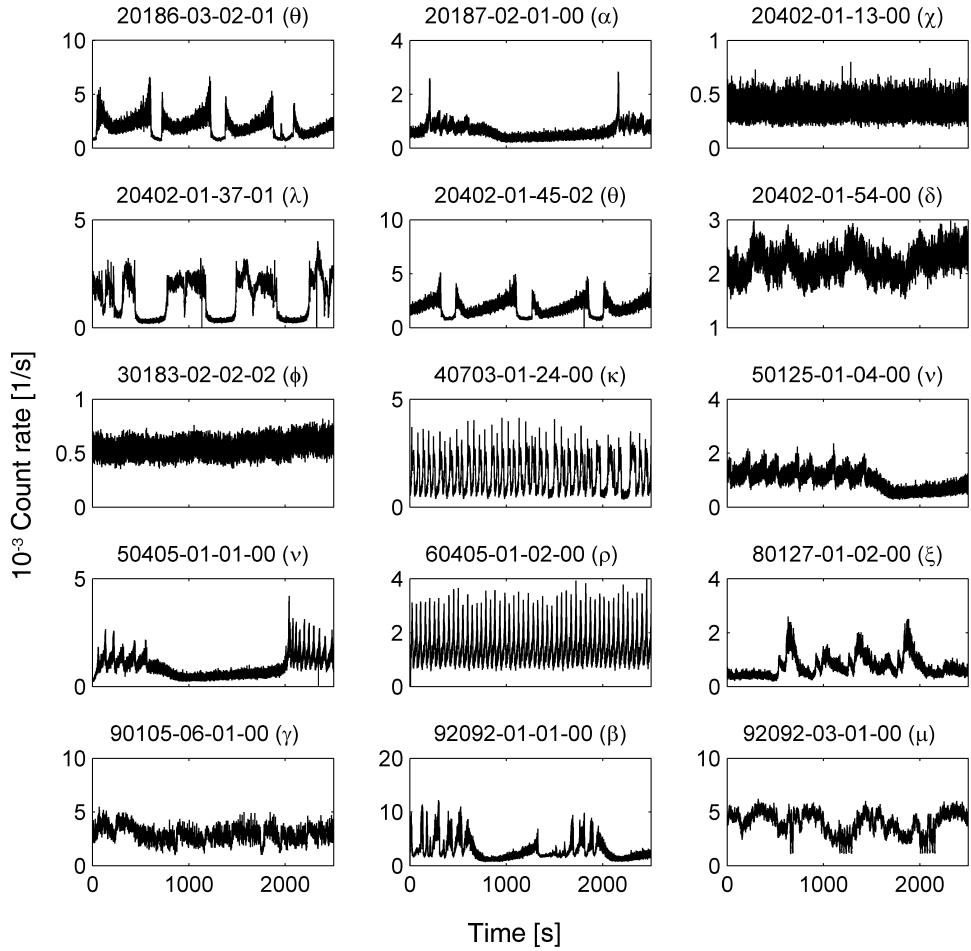}
\caption{First 2,500-second record of continuous activity for light curves of 15 sample observations (soft X-ray energy band).\label{fig1}}
\end{center}
\end{figure}

In order to sample the stochastic variability in GRS 1915+105, we analyze 15 of the observations of 
this system (13 different X-ray states) made with the Rossi X-ray Timing Explorer over the last 15 years, which are illustrated in Figure\ \ref{fig1}. Each observation consists of several $\sim2,500$-second intervals separated by gaps. The specific observations are selected to study at least one long instance of each of GRS 1915+105's known states (except for the $\omega$ state, for which we did not find a long enough continuous data set). From each observation, we extract X-ray lightcurves in two energy bands (soft: 2-6 keV, and hard: 15-60 keV) with sampling frequencies $f_d$ of 16 and 128 Hz. We also extract X-ray background lightcurves in both energy bands to subtract from each time bin, and we correct the lightcurve count rates for detector dead time. Taking into consideration that all analyzed time series exhibit Poisson noise in the highest frequency range, we add a term to our interpolations  (\ref{eq5}-\ref{eq9}) accounting for this noise. For the relatively low frequencies considered here, the expression for the dead-time-corrected Poisson noise level \citep{Zha95,MRG97} can be approximated by a constant value (in general, it is frequency-dependent; see Appendix C). Considering only continuous 2,500-s intervals in two energy bands, we analyze 94 time series in total. 

We perform the analysis at $T_M = T/9 \approx 280$ s to make sure the autocorrelation-based power spectrum estimate is close to the actual power spectrum value. The spectral FNS parameters are also found for the power spectrum estimate calculated by the Welch method \citep{Wel67} (averaging 8 windows of size $T/9$ with a 50\% overlap). The Welch method is similar to methods for computing power spectra commonly used in X-ray astrophysics \citep{MRG97}. The difference between the parameter values found by the two methods rarely exceeded $10\%$, which implies that the autocorrelation-based power spectrum estimate is adequate for our analysis.

For the 15 observations analyzed in this study, the variations of FNS parameters within a given observation were not significant in most cases (generally less than $10\%$), which suggests that the parameterization method is reliable and that specific variability modes last longer than the observations studied here. Table 1 lists the observation-averaged values of parameters for the 12 observations where the contribution of Poisson noise is significant in the frequency range close to 16Hz. The other 3 observations, which have a relatively high signal-to-Poisson-noise ratio in that frequency range, are parameterized for a higher sampling frequency $f_d=128$ Hz (Table 2). 
The parameters $S_s(T_{0}^{-1})$, $S_s(T_{01}^{-1})$, $S_s(T_{02}^{-1})$, $\sigma$, $\sigma_1$, and $\sigma_2$ are not listed in the tables because their values cannot be reliably determined when a significant Poisson noise level is present. 

\begin{landscape}

\begin{table}[!t]
\renewcommand{\arraystretch}{1.1}
\footnotesize
\caption{Parameters for the light curves of 12 GRS 1915+105 observations with 1S, 1F, or RN behavior ($f_d = 16$ Hz)}
\label{table1}
\begin{tabular}{ccc |ccccccc|ccccccc}
\tableline
\tableline
\multicolumn{3}{c|}{} & \multicolumn{7}{c|}{Soft X-rays} & \multicolumn{7}{c}{Hard X-rays}  \\
\tableline
Observation ID&Class&{$N_{ci}$\tablenotemark{1}}&$\mu_V$\tablenotemark{2}, 1/s&Type&PN\%\tablenotemark{3}&$n$&$H$&$T_0$, s&$T_1$, s&$\mu_V$\tablenotemark{2}, 1/s&Type&PN\%\tablenotemark{3}&$n$&$H$&$T_0$, s&$T_1$, s\\
\tableline
90105-06-01-00&$\gamma$&4&2790&1S&15&1.73&0.37&13&17&122&1S&30&2.15&0.59&5&6\\

40703-01-24-00&$\kappa$&3&1443&1S&20&2.55&0.77&5&5&89&1S&35&2.61&0.86&3&3\\

60405-01-02-00&$\rho$&3&1548&1S&15&2.15&0.58&6&7&100&1S&35&1.91&0.47&2&3\\

20402-01-54-00&$\delta$&3&2134&1F&15&1.2&0.1&--&--&160&1F&50&1.04&0.06&--&--\\

20402-01-37-01&$\lambda$&2&1409&1F&25&1.91&0.46&--&--&182&1F&30&1.57&0.29&--&--\\

92092-03-01-00&$\mu$&4&3814&1F&20&1.7&0.35&--&--&148&1F&45&1.94&0.48&--&--\\

80127-01-02-00&$\xi$&4&889&1F&10&1.47&0.25&--&--&44&1F&65&3.06&0.92&--&--\\

30183-02-02-02&$\phi$&3&611&1F&40&1.36&0.21&--&--&49&1F&70&1.88&0.3&--&--\\

20187-02-01-00&$\alpha$&2&581&1S&10&1.53&0.27&50&86&129&RN&--&--&--&--&--\\

50125-01-04-00&$\nu$&3&1102&1F&5&1.29&0.16&--&--&107&RN&--&--&--&--&--\\

50405-01-01-00&$\nu$&2&848&1F&5&1.37&0.18&--&--&92&RN&--&--&--&--&--\\

20402-01-13-00&$\chi$&3&387&RN&--&--&--&--&--&119&RN&--&--&--&--&--\\

\tableline
\end{tabular}
\tablenotetext{1}{number of intervals in observation}
\tablenotetext{2}{average count rate}
\tablenotetext{3}{percentage of Poisson noise in the highest-frequency range - determined for the plot in double logarithmic scale}
\end{table} 

\begin{table}[!t]
\renewcommand{\arraystretch}{1.1}
\footnotesize
\caption{Parameters for the light curves of 3 GRS 1915+105 observations with 2S behavior ($f_d = 128$ Hz; notes same as in Table 1)}
\label{table2}
\begin{tabular}{ccccccccccccccc}
\tableline
\tableline
Observation Id&Class&$N_{ci}$&Band&$\mu_V$,1/s&Type&$n_1$&$H_1$&$T_{01}$,s&$T_{11}$,s&$n_2$&$H_2$&$T_{02}$,s&$T_{12}$,s\\
\tableline
92092-01-01-00 & $\beta$ &3&Soft&1604&2S&1.73&0.53&0.22&0.20&3.07&1.06&12&10\\
&&&Hard&121&2S&1.76&0.58&0.16&0.14&2.49&0.66&30&36\\
\tableline
20186-03-02-01 & $\theta$ &4&Soft&1738&2S&2.53&0.88&0.13&0.13&2.3&0.65&34&42\\
&&&Hard&228&2S&2.19&0.75&0.11&0.10&2.5&0.7&38&46\\
\tableline
20402-01-45-02 & $\theta$ &4&Soft&1783&2S&2.38&0.82&0.13&0.13&2.21&0.58&39&56\\
&&&Hard&232&2S&1.9&0.66&0.13&0.11&2.68&0.74&35&44\\
\tableline
\end{tabular}
\end{table} 

\end{landscape}

\begin{figure}
\figurenum{2}
\begin{center}
\includegraphics[width=14cm]{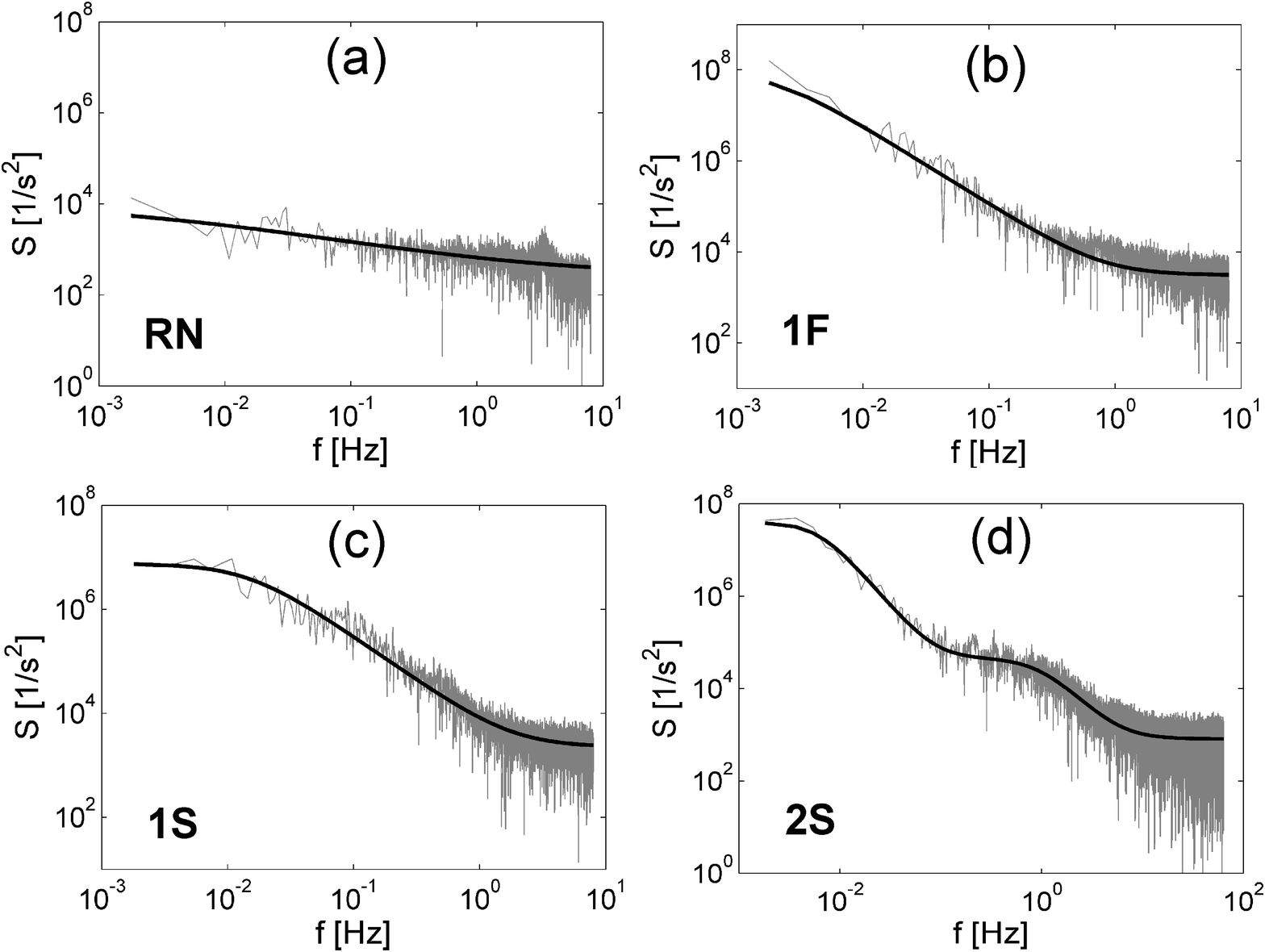}
\caption{\label{fig2} Typical power spectrum plots for each mode of stochastic variability [thick black line is spectral FNS interpolation; thin gray line is experimental power spectrum estimate]: (a) RN, class $\chi$, observation 20402-01-13-00, 1$^{\textrm{st}}$ interval, soft X-ray energy band, $n \approx 0.52$; (b) 1F, class $\mu$, observation 920920-03-01-00, 3$^{\textrm{rd}}$ interval, soft X-ray energy band; (c) 1S, class $\gamma$, observation 90105-06-01-00, 3$^{\textrm{rd}}$ interval, soft X-ray energy band; (d) 2S, class $\theta$, observation 20402-01-45-02, 4$^{\textrm{th}}$ interval, soft X-ray energy band.}
\end{center}
\end{figure}

\begin{figure}
\figurenum{3}
\begin{center}
\includegraphics[width=14cm]{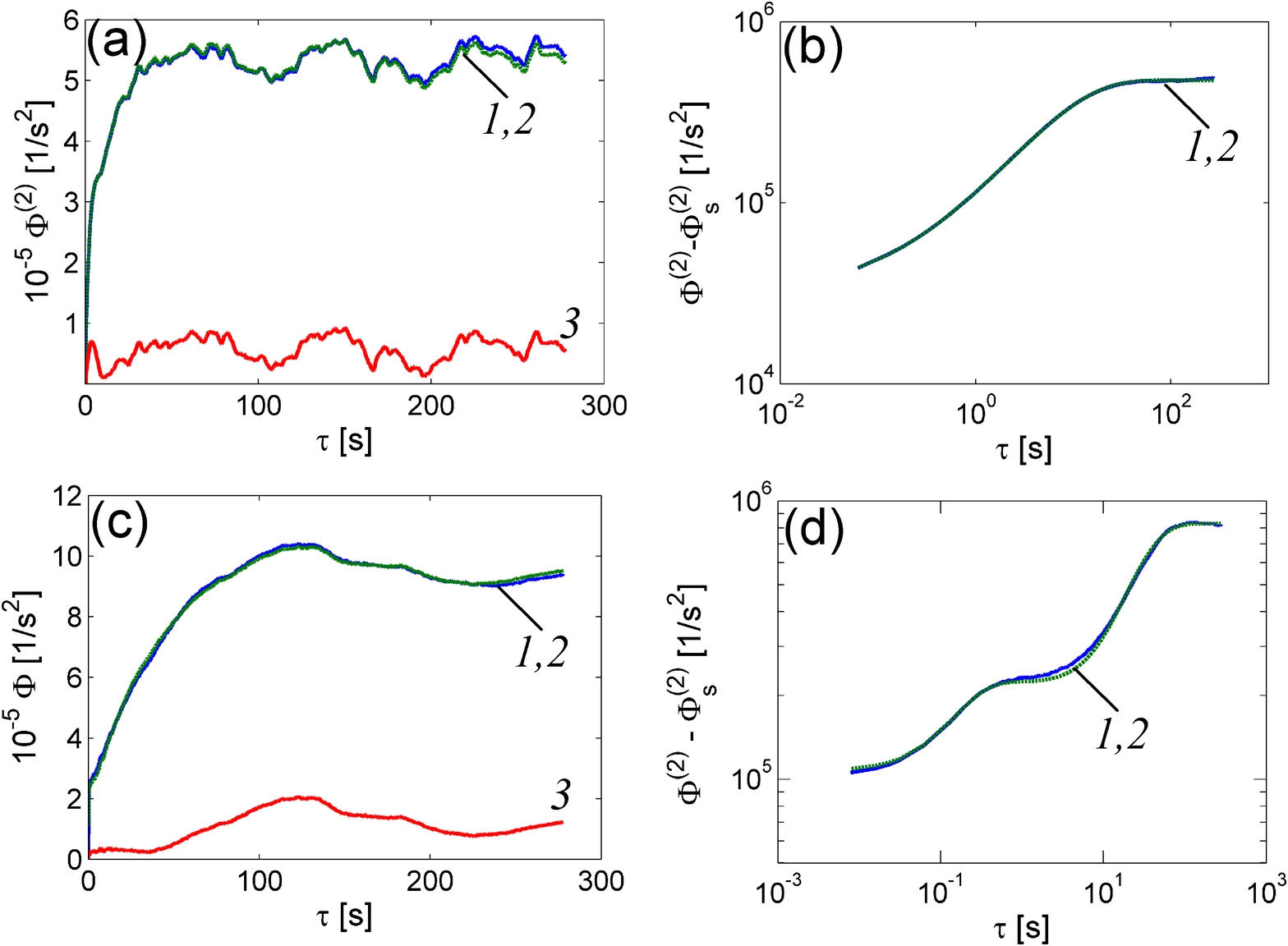}
\caption{\label{fig3} Sample structure function plots for variability modes 1S and 2S [1 (blue)- experimental curve, 2 (green) - interpolation, 3 (red) - resonant component of structure function]: (a) structure function plot for 1S, class $\gamma$, observation 90105-06-01-00, 3$^{\textrm{rd}}$ interval, soft X-ray energy band and (b) its stochastic component of structure function; (c) structure function plot for 2S, class $\theta$, observation 20402-01-45-02, 4$^{\textrm{th}}$ interval, soft X-ray energy band and (d) its stochastic component of structure function. See the electronic edition of the Journal for a color version of this figure.}
\end{center}
\end{figure}

Our analysis shows that the noise behavior for the observations considered in this study exhibits four different modes of stochastic variability: ``random" (RN), power-law (1F), one-scale (1S), and two-scale (2S). The RN variability, which is illustrated in Fig \ref{fig2}(a), corresponds to values of $n$ less than unity and the overall variation of spectral power in the considered frequency range by less than two orders of magnitude. The 1F variability is characterized by a power-law ($1/f^n$) dependence at low and intermediate frequencies and a flat Poisson noise level at highest frequencies, as can be seen in Figure \ref{fig2}(b). This implies that the stochastic process for this variability mode exhibits a form of self-similar behavior and is far from reaching the steady state (characteristic time scale) at the averaging interval and frequency range considered here. The 1S variability follows the interpolation (\ref{eq5}) at low and intermediate frequency ranges and exhibits a Poisson noise at highest frequencies, as illustrated in Figure \ref{fig2}(c). The 2S variability is adequately described by interpolation (\ref{eq7}) and contains two characteristic time scales, as can be seen in Figure \ref{fig2}(d). In Figure\ 3, we present the structure functions (panels (a) and (c)) and their stochastic components (panels (b) and (d)) for sample 1S and 2S variability modes, respectively. The stochastic component is obtained by subtracting the resonant component from the overall structure function. For the 1S mode, the stochastic component is described by an anomalous diffusion process reaching a steady state \citep[Figure\ \ref{fig3}(b); ][]{Tim10a}, while for the 2S case a two-scale process is implied (Figure\ \ref{fig3}(d)).


It can be seen that the parameters for two instances of the same class ($\nu$ in Table 1 and $\theta$ in Table 2) stay approximately same, which suggests that a particular stochastic variability (certain variability mode and ranges for parameter values) may characterize a class of observations. However, an exhaustive analysis of all X-ray emission observations for GRS 1915+105 is needed to draw definite conclusions in this context.

For most of analyzed intervals corresponding to variability modes 1F, 1S, and 2S, we noticed that $ n \approx 2 H + 1$ ($ n_1 \approx 2 H_1 + 1$, $ n_2 \approx 2 H_2 + 1$), as expected for fractional Brownian motions \citep{Mal99}. We also observed that the characteristic time scales in power spectrum estimate and structure function were close to each other for most of the intervals characterized by variability modes 1S and 2S. Both of these facts imply that the random-walk component plays the dominant role in the stochastic variability of X-ray emission, and the contribution of highest-frequency inertial spike-irregularities in these data is not significant \citep{Tim07a}. This can most likely be attributed to the ``erasing" effect of Poisson noise at highest frequencies, which is illustrated in Figure \ref{fig2} and by the values of parameter PN\% in Table 1.

Tables 1 and 2 as well as interval-specific analysis suggest that there is no statistically significant difference between the FNS parameter values for soft and hard X-ray energy bands in the same interval. The only differences are in the count rate magnitude (it is smaller for hard X-rays) and the frequency range dominated by the Poisson noise (PN$\%$ is higher for hard X-rays). Soft X-ray emission is nominally associated with the thermal spectrum of the accretion disk around the black hole, while hard X-ray emission is generally believed to be related to Compton scattering in a hot electron corona around the disk. The fact that the FNS parameters for these energy bands are close in value could suggest that at these frequencies, there is a common mechanism producing stochastic variability. Future analysis in more energy bands could help determine whether the FNS parameters are truly energy-independent.

Let us compare the classification in Tables 1 and 2 with the results obtained by the nonlinear time series analysis of total count rates for sample observations corresponding to 12 different states \citep[][Table 1]{Har11}. As total count rates are dominated by the soft X-ray energy band, we compare the results of \citet{Har11} with the soft X-ray classification presented in this paper. The noise variability for class $\chi$ is characterized as random by both analyses. Class $\phi$ is marked as random by \citet{Har11}, but is assigned to 1F here. The difference, in our opinion, is due to the significant effect of the Poisson noise within half of the considered frequency range (high values of PN$\%$), which was not separately taken into account in the nonlinear time series analysis. Class $\gamma$ was marked as random by \citet{Har11}, but is categorized as 1S by FNS. This discrepancy could be due to the plateau in the low-frequency range of the power spectrum \citep[Figure\ \ref{fig2} in our study;][Figure\ 10]{Har11}: although it may mimic white noise in the nonlinear time series analysis, in FNS the plateau corresponds to the loss of correlations (at $\tau > T_0$) within a sequence of stochastic irregularities. Class $\delta$ is categorized as random by \citet{Har11}, but is marked as 1F in FNS. Considering that the value of $n$ is close to unity in this case (boundary range), additional sample observations for this class would be needed to determine which method is more accurate. Classes $\kappa$, $\lambda$, and $\mu$ were labeled by \citet{Har11} as ``deterministic nonlinear + colored noise". All three are marked as 1F or 1S in this study, which corresponds to a variant of ``colored noise". In classes where there are strong periodic or quasi-regular (``resonant") components (e.g.\  $\alpha$, $\beta$,  $\theta$, $\nu$, $\kappa,$ $\lambda$, and $\rho$), it is difficult to compare our categorizations directly to the results of \citet{Mis06,Har11}, because in contrast to those studies, we subtract the resonant components prior to performing our analysis. In other words, the regular (deterministic) behavior observed in SVD plots \citep{Mis06,Har11} could be dominated by the resonant periodicities, rather than the stochastic component studied here. Moreover, classes $\beta$ and $\theta$ contain two separate stochastic scales in different frequency ranges, which would be difficult to capture in terms of chaos theory, originally developed for systems of nonlinear ordinary differential equations with one characteristic scale.

The above analysis demonstrates that the random-walk component plays the dominant role in the stochastic variability of modes 1F, 1S, and 2S. The random-walk component can be interpreted in terms of a dissipative process of anomalous diffusion \citep{Tim10a}. It has been argued  that stochastic fluctuations of the viscosity in the accretion flow may be responsible for X-ray variability on a wide range of time scales, as well as the presence of multiple characteristic time scales \citep{L97,T07,Wil09,Utt11}. This is because variations in the viscosity lead to changes in the accretion rate, which propagate towards the black hole, resulting in flickering X-ray emission. In principle, the radial dependence of the amplitude of the viscous fluctuations can determine the slope of the power spectrum, and different components in the accretion flow (e.g.\ corona, inner accretion disk, and outer accretion disk) can contribute at different frequencies. The stochastic modes described here may  be related to these viscous fluctuations in the accretion disk, in which case a more comprehensive analysis with FNS could provide a useful characterization of  X-ray states from the perspective of stochastic variability. 

\section{CONCLUDING REMARKS}

Our analysis shows that the consideration of different frequency ranges in flicker-noise spectroscopy allows one to analyze the stochastic variability in complex signals of GRS 1915+105, despite the significant contribution of Poisson noise in the highest frequency range, high low-frequency quasi-regular variability, and presence of multiple stochastic time scales. As a proof of concept for future study, the examination of sample observations for 13 different states demonstrates that there are at least four types of stochastic variability in the X-ray emission of GRS 1915+105: random (RN), power-scale (1F), one-scale (1S), and two-scale (2S). The last three are related to random walk processes interpretable in terms of a dissipative process of anomalous diffusion. In this case, the random walk processes could be related to viscosity fluctuations in the accretion disk.

Other stochastic variability modes are also possible in the frequency ranges under consideration. For example, it is natural to expect a stochastic variability mode representing an intermediate category between 1S and 2S, i.e., a superposition of one full scale (1S) and self-similar (1F) behavior in different frequency ranges. This is borne out by astrophysical observations, which exhibit a wide range of extremely complex power spectra, and are frequently modeled as the sum of a number of different components \citep{Bel02,K06}. An exhaustive analysis of all available X-ray emission observations for GRS 1915+105 is needed to get a complete list of variability modes, obtain more accurate values of FNS parameters for each known class of the binary system, and revisit the established classification from the stochastic perspective. A similar analysis of other accreting black holes could help to identify the effect of the extreme variability of GRS 1915+105 on the FNS parameters, and to determine the true relationship of these parameters to stochastic processes in the disk. It would also be interesting to study the variability at other wavelengths, e.g.,\ in observations of the radio jet with the Ryle Telescope \citep{PF97,Fender97} or the \textit{EVLA}, via FNS parameterization and cross-correlation with X-rays, in order to compare stochastic variability in different accretion processes. 

\acknowledgments

J.N. was supported in part by the Harvard University Graduate School of Arts and Sciences, Chandra grant AR0-11004X, and the National Aeronautics and Space Administration through the Smithsonian Astrophysical Observatory contract SV3-73016 to MIT for support of the Chandra X-ray Center, which is operated by the Smithsonian Astrophysical Observatory for and on behalf of the National Aeronautics Space Administration under contract NAS8-03060.

\appendix

\section{Parameterization algorithm for one-scale case in discrete form}
\label{A}

 Consider a time series $V_d(k)$. The subscript $d$ here and below is used to denote the discrete form of expressions. Let $N_t$ be the number of points corresponding to the selected averaging interval $T$, $M$  be the number of 
 points used in estimating the autocorrelation function. In this case, the parameterization procedure can be written as follows:
\newline 1. Calculate the arithmetic mean for the signal:
\begin{equation}
\mu _V  = {1 \over {N_t }}\sum\limits_{k = 1}^{N_t} {V_d\left( k \right)}. \label{eqA7}
\end{equation}
\newline 2. Subtract the arithmetic mean from the series $V_d(k)$:
\begin{equation}
\mathop {V_d}\limits^-  \left( k \right) = V_d\left( k \right) - \mu _V. \label{eqA8}
\end{equation}
\newline 3. Calculate the autocorrelation function for the series $\mathop {V_d}\limits^ -$:
\begin{equation}
 \psi_d (p) = {1 \over {N_t  - p}}\sum\limits_{k = 1}^{N_t  - p} {\mathop {V_d}\limits^ -  \left( k \right)\,\mathop {V_d}\limits^ -  \left( {k + p} 
\right)},   p = 0..M.  \label{eqA9}
\end{equation}
The autocorrelation interval $M$ should not be higher than $N_t/4$ (higher values of $M$ will result in 
the loss of statistical information in estimating the autocorrelation function). To go from discrete form to the continuous one, one can use the following expression: $p = N_t \tau/T$.
\newline 4. Calculate the discrete cosine transform of the autocorrelation function:
\begin{equation}
 S_d(q) = \psi_d (0) + \psi_d \left( M \right)\left( { - 1} \right)^q  + 2\sum\limits_{p = 1}^{M - 1} {\psi_d (p)\cos \left( {{{\pi  \,q\,p} \over 
M}} \right)}, \label{eqA10}
\end{equation}
 where $q=0..M$. For $q=1..M-1$, $S_d(q)$ should be multiplied by 2, which is the standard procedure for discrete Fourier 
transforms to take into account the spectral values in the second half of the frequency range. Here, relations $q=2 f f_d^{-1} M$ and $S_d(q)=S(f)\times f_d$  describe the equivalence between the discrete and continuous 
 forms of power spectrum estimate.
 \newline 5. Calculate $S_{sd}(0)$  as the average value of the power spectrum for the points 1 and 2 (point 0, which corresponds to the zero 
frequency, is not used in calculating $S_{sd}(0)$):
\begin{equation}
S_{sd} \left( 0 \right) = {{S_d\left( 1 \right) + S_d\left( 2 \right)} \over 2}. \label{eqA11}
\end{equation}
\newline 6. Interpolate $|S_d(q)|$ given by Equation~(\ref{eqA10}) using the expression:
\begin{equation}
S_{sd} (q) = {{S_{sd} (0)} \over {1 + (\pi {q \over M}T_{0d} )^n }} + P_S \label{eqA12}
\end{equation}
 by the method of nonlinear least-square fitting to determine the values of parameters $n$ and $T_{0d}$. The constant term $P_S$ is an estimate of the Poisson noise level calculated as the average value of $S_d$ for 100 highest-frequency points. The fitting is done on the basis of a 
 double logarithmic scale, dividing the entire series into a set of equal intervals (in the calculations presented in this study we took 200 equal intervals). We used the trust-region algorithm for nonlinear square 
fitting, which is built in MATLAB v.7 or higher \citep{Bra99}.
\newline 7. Separate out the resonant component:
\begin{equation}
S_{rd} \left( q \right) = S_d\left( q \right) - S_{sd} \left( q \right), q = 0..M. \label{eqA13}
\end{equation}
 \newline 8. Calculate the autocorrelation function for the resonant component as the inverse discrete cosine transform of $S_{rd}(q)$. When 
 $q=1..M-1$, divide  $S_{rd}(q)$ by 2 to take into account the spectral values in the second half of the frequency range. Then calculate the 
inverse cosine transform:
\begin{eqnarray}
\psi_{rd} (p) = {1 \over {2M}} \left\{ {S_{rd} (0) + S_{rd} \left( M \right)\left( { - 1} \right)^p} \right\} \nonumber \\
+ {1 \over {2M}} \left\{ {2\sum\limits_{q = 1}^{M - 1} {S_{rd} (q)\cos \left( {{{\pi  \,p\,q} \over M}} \right) } } \right\}. \label{eqA14}
\end{eqnarray}
9. Calculate the difference moment for the resonant component:
\begin{equation}
\Phi_{rd}^{(2)} (p) = 2\left[ {\psi_{rd} (0) - \psi_{rd}  (p)} \right], p = 0..M. \label{eqA15}
\end{equation}
The continuous equivalent of $\Phi_{rd}^{(2)} (p)$ is $\Phi_{r}^{(2)} (\tau)$.
\newline 10. Calculate the difference moment for the experimental series:
\begin{equation}
 \Phi_d^{(2)} (p\,) = {1 \over {N_t  - p}}\sum\limits_{k = 1}^{N_t  - p} {\left[ {\mathop {V_d}\limits^ -  (k) - \mathop {V_d}\limits^ -  (k + p)} 
\right]^2 }. \label{eqA16}
\end{equation}
\newline 11. Calculate the difference moment for the random component:
\begin{equation}
\Phi _{esd}^{(2)} (p) = \Phi_d ^{\left( 2 \right)} (p) - \Phi _{rd}^{(2)} (p). \label{eqA17}
\end{equation}
The continuous equivalent of $\Phi_{esd}^{(2)} (p)$ is $\Phi_{es}^{(2)} (\tau)$.
 \newline 12. Determine the parameters $\sigma$, $H$, $T_{1d}$, $P_\Phi$ by fitting $\Phi _{esd}^{(2)} (p)$ in Equation~(\ref{eqA17}) to the interpolation 
expression of the anomalous diffusion type \citep{Tim10a}:
\begin{equation}
 \Phi_{sd}^{(2)} (p) = 
2\sigma ^2  \times \left[ {1 - \Gamma ^{ - 1} (H) \times \Gamma (H ,p/T_{1d} )} \right]^{2} + P_\Phi, \label{eqA18}
\end{equation}
 where $\Gamma (s,x) = \int\limits_x^\infty  {\exp ( - t) t^{s - 1} dt},\, \Gamma (s) = \Gamma (s,0)$, using the same least-square fitting 
method as in step 6. Here, the constant term $P_\Phi$ is an estimate of the Poisson noise contribution to the structure function.
 \newline 13. Calculate $S_{sd}(T_{01d}^{-1})$ by Equation~(\ref{eqA12}).
 \newline 14. After the values of all six FNS parameters - $\sigma$, $T_{0d}$, $T_{1d}$, $H$, $n$, $S_{sd}(T_{0d}^{-1})$ - are determined, calculate the dimensional values for $T_{0d}$, $T_{1d}$,  $S_{sd}(T_{0d}^{-1})$: $T_0 = T_{0d} \times \Delta t$, $T_1 = T_{1d} \times \Delta t$, $S_s(T_{0}^{-1}) = S_{sd}(T_{0d}^{-1}) \times \Delta t$, where $\Delta t = f_d^{-1}$ .
\newline 15. Calculate the relative error $\epsilon_\Phi$ in the interpolation of difference moment $\Phi_d^{(2)} (p)$:
\begin{equation}
 \epsilon _\Phi   = {{\sum\limits_{p = 1}^M {\left| {\Phi_d ^{(2)} \left( p \right) - \Phi _{rd}^{(2)} \left( p \right) - \Phi _{sd}^{(2)} \left( p 
\right)} \right|} } \over {\sum\limits_{p = 1}^M {\Phi_d ^{(2)} \left( p \right)} }} \times 100\%.  \label{eqA19}
\end{equation}
 Here, the error is determined as the ratio of the difference of areas between the experimental structure function and the total 
 interpolation function to the area of the experimental structure function. The areas are calculated by numerical integration using the 
rectangle method because the original series have a rather large number of points. The parameterization is successful if 
$\epsilon _\Phi   \le 10\% $ \citep{Tim10a}.

\section{Parameterization algorithm for two-scale case in discrete form}
\label{B}

Steps 1 through 5 are same as in the one-scale case (Appendix A). 

In step 6, the following interpolation is used

\begin{equation}
S_{sd} (q) = {{{S_{s1d} (0)} \over {1 + (\pi {q \over M} T_{01d} )^{n1} }} +{ {S_{sd}(0)-S_{s1d} (0)} \over {1 + (\pi {q \over M} T_{02d} )^{n2} }}} + P_S.
\label{eqB1}
\end{equation}
Here, the method of nonlinear least-square fitting is applied in log-log scale to determine the values of $S_{s1d}(0)$, $T_{01d}$, $n_1$, $T_{02d}$, and  $n_2$.

Steps 7 through 11 are same as in the one-scale case (Appendix A). 

In step 12, we first use the following interpolation to estimate the values of parameters for the first scale:

\begin{eqnarray}
 \Phi_{sd}^{(2)} (p) = 
2\sigma_1^2  \times [ {1 -  { \Gamma (H_1 ,p/T_{11d} ) \over \Gamma (H_1 )}}  \nonumber \\
+ { \sigma_2 \over \sigma_1 }  \left({1 - {\Gamma (H_2 ,p/T_{12d} ) \over \Gamma (H_2 )}}\right)   ]^{2} + P_\Phi, 
\label{eqB2}
\end{eqnarray}

The nonlinear least-square fitting method in log-log scale is applied to find the values of parameters $\sigma_1$, $H_1$, $T_{11d}$, $\sigma_2$, $H_2$, and $T_{12d}$. The parameter $P_\Phi$ is estimated as the value of $\Phi^{(2)}_{esd}(p)$ at $ p = 1$.  The set corresponding to the lower value of $T_{11d}$ and $T_{12d}$ is used as the values for $\sigma_1$, $H_1$, and $T_{11d}$.

The parameter $\tau_{0d}$ is estimated as the value of $p$ after which $\Phi^{(2)}_{esd}(p) - P_\Phi$ exceeds $2 \sigma_1^2$ by at least $3\%$. Then we use interpolation
\begin{equation}
 \Phi_{sd}^{(2)} (p) = 
2\sigma_1^2  +  2 \sigma_2^2   \left[ \left({1 - {\Gamma (H_2 ,{{p-\tau_{0d}} \over {T_{12d}}} ) \over \Gamma (H_2 )}}\right)  \right ]^{2} + P_\Phi
\label{eqB3}
\end{equation}
for $p > \tau_{0d}$ to find the values of $\sigma_2$, $H_2$, and $T_{12d}$.

At step 13, calculate $S_{sd}(T_{01d}^{-1})$ and $S_{sd}(T_{02d}^{-1})$ by Equation~(\ref{eqB1}).

At step 14, calculate the dimensional values for $T_{01d}$, $T_{02d}$, $T_{11d}$, $T_{12d}$,  $S_{sd}(T_{01d}^{-1})$, $S_{sd}(T_{02d}^{-1})$: $T_{01} = T_{01d} \times \Delta t$, $T_{02} = T_{02d} \times \Delta t$, $T_{11} = T_{11d} \times \Delta t$, $T_{12} = T_{12d} \times \Delta t$,  $S_s(T_{01}^{-1}) = S_{sd}(T_{01d}^{-1}) \times \Delta t$, $S_s(T_{02}^{-1}) = S_{sd}(T_{02d}^{-1}) \times \Delta t$, where $\Delta t = f_d^{-1}$ .

Step 15 is same as in the one-scale case (Appendix A). 

\section{Approximation for dead-time-corrected Poisson noise level expression}
\label{C}

The formula for computing the Poisson power corrected for instrument dead time is given in Refs. \citep{Zha95,MRG97}. Adapting this for power spectra based on the autocorrelator, we have:

\begin{eqnarray}
S_P(f) = A \times(2 \left [1-2 r_0 \tau_d \left( 1- {\tau_d \over {2 \tau_b}} \right) \right] \nonumber  \\
- 2 { (N_t - 1) \over N_t} r_0 \tau_d \left( {\tau_d \over t_b} \right) \cos\left(2 \pi t_b f \right) \nonumber \\
+2 r_{vle} r_0 \tau_{vle}^2 \left[ {{\sin \left(\pi \tau_{vle} f \right)} \over {\pi \tau_{vle} f }}\right]^{2} ),
\label{eqC1}
\end{eqnarray}
where $A$ is a scale factor between FNS units and standard astrophysical power spectra (where pure Poisson noise has $S_P(f)=2$), $r_0$ is the count rate per proportional counter unit (PCU), $N_t$ is the number of frequency points, $r_{vle}$ is the rate of very large events (e.g.\ cosmic rays; VLE) per PCU, $t_b$ is the bin size, $\tau_d$ is the dead time per event, $\tau_{vle}$ is the dead time window for each VLE. In our case, $r_{vle}=200$ counts/s/PCU, $t_b$=0.0625 or 0.0078 s, $\tau_d = 10 \mu$s, and $\tau_{vle} = 150 \mu$s.

It is more convenient for analysis to write Equation (\ref{eqC1}) as 

\begin{equation}
S_P(f) =   B + C \cos \left[ D f \right] + E ,
\label{eqC2}
\end{equation}
where 

$B = 2 A \left [1-2 r_0 \tau_d \left( 1- {\tau_d \over {2 \tau_b}} \right) \right]$, 

$C = - 2 A { (N_t - 1) \over N_t} r_0 \tau_d \left( {\tau_d \over t_b} \right)$, 

$D = 2 \pi t_b$, 

$E = 2 A r_{vle} r_0 \tau_{vle}^2 \left[ {{\sin \left(\pi \tau_{vle} f \right)} \over {\pi \tau_{vle} f }}\right]^2 \approx  2 A r_{vle} r_0 \tau_{vle}^2 $.

It can be easily shown for our data that $ B \gg C + E$ (usually by at least 2 orders of magnitude). This implies that in our study we can use the constant-value approximation

\begin{equation}
S_P(f) \approx B.
\label{eqC3}
\end{equation}

\bibliography{AJ_FNS} 
 
\end{document}